\documentclass[12pt]{article}
\usepackage{amscd,amsmath,amssymb,amsfonts,
         latexsym,  verbatim}
\newtheorem{theorem}{Theorem}[section]

\begin{document}

\begin{center}
 {\bf   Convergent and Anti-diffusive Properties of Mean-Shift  Method}
 \end{center}

\begin{center}
{ Xiaogang Wang and Jianhong Wu\\
Department of Mathematics and Statistics\\
York University, Toronto, Canada}
\end{center}

\begin{abstract}
An analytic   framework based on partial differential equations is derived for certain dynamic clustering methods.
 The proposed mathematical framework is based on the application of the conservation law  in physics to characterize  successive transformations of the underlying probability density function.
It  is then applied  to analyze the convergence and stability of mean shift type of dynamic clustering algorithms.
Theoretical analysis shows that un-supervised mean-shift type of algorithm is intrinsically unstable.
It is proved that the only possibility of a correct convergence for unsupervised mean shift type of algorithm is  to  transform the original probability density  into a   multivariate normal distribution with no dependence struture.
Our analytical results suggest that a more stable and convergent mean shift algorithm  might be achieved
by adopting a judiciously  chosen supervision mechanism.
\end{abstract}


\section{Introduction}
\label{sect:Intro}
Clustering is a  process of partitioning a set of objects into subgroups according to a certain measure of similarity. Cluster analysis has many applications in mining large data sets arising from the study of biology and climate, which normally need to be partitioned into much smaller and homogeneous groups. Hastie et al. (2001)  and Han and Kamber (2006) both offer excellent reviews on  clustering algorithms with different emphasis, and  Gan, Ma and Wu (2007) provides additional discussions.

For many classical clustering algorithms, such as K-means (MacQueen 1967; Hartigan and Wong 1979) and PAM (Kaufman and Rousseeuw 1990), the number of clusters or sub-populations needs to be specified by the user. A popular approach to select the number of clusters is to optimize a certain measure of strength of the found clusters (Tibshirani {\it et. al} 2000 and Fraley and Raftery 2002).  One alternative is to first partition the data into many small clusters, and then merge these small clusters until no cluster can be merged (Frigui and Krishnapuram 1999), while another alternative is to extract one cluster at a time (Zhung et al. 1996).  The determination of the number of clusters, however, remains to be a challenging problem when the clusters assume non-normal shapes with blurring or even slightly overlapping boundaries. Another difficult clustering issue is to specify the exact functional form of the underlying probability density functions. For convenience and ease of computations, Gaussian or normal distributions are often employed. However, any assumed density would introduce distortions of relative positions of data points in a high dimensional manifold  when the assumed probability distribution deviates significantly from the underlying one which is usually unknown.
Non-parametric   clustering algorithms have emerged from various disciplines in the last twenty years with a wide range of applications
including pattern recognition and image analysis, see Han and Kamber (2006) and Gan {\it et al.} (2007).

In this paper, we are concerned with dynamic clustering algorithms.
One common theme among several dynamic clustering algorithms is the employment of a   gravitational field. To be more specific, this type of dynamic clustering methods treats data points as autonomous agents or particles and iteratively move them towards cluster centers or focal points. One typical approach, including the gravitational clustering algorithm (Wright 1977; Kundu 1999; Sato 2000; Wang and Rau 2001), considers each data point as a particle of unit mass with zero velocity which is gradually moved towards a cluster center following a gravitational law. The theoretical (dynamical) properties of these gravitational clustering algorithms, however, have not been fully understood  thirty years after the original idea was introduced.

The most famous and representative algorithm under the category of dynamic non-parametric clustering is arguably the so-called {\it mean-shift} algorithm and its variations proposed by Fukunaga and Hostetler (1975), Cheng (1995), Comaniciu and Meer,(2002).  It has received increasing attentions in the literature due to its flexible and adaptive nature.
Given a kernel function $K$ and a weight function $w$,  the generalized mean-shift operation is given by
\begin{equation}
T(x) = \frac{\sum K(x, s) \;w(s)\, s} {\sum K(x, s) \;w(s)}.
\end{equation}
This operation originates from the intuitive idea of finding the gradient in kernel density estimation since data points are transformed toward denser regions by using a function of the kernel estimate of the density function. There are many variations and improvements of this algorithm proposed in the literature, including for example, Comaniciu and Meer,(2002), Virmajoki (2002),  Shi et al. (2005) and Wang et al. (2007a).  In the data sharpening procedure proposed by Choi and Hall (1999), originally designed to reduce bias by pushing data points at the boundary a bit closer to the center, the movement of the data points also resembles the one proposed in the mean-shift method. Woolfold and Braun (2006) apply the data sharpening method for the identification and tracking of spatial temporal centers of lightning activity. Comaniciu and Meer (2002) employed the mean shift as the core procedure for their robust pattern recognition method and achieved excellent performance for image analysis and segmentation.

Although the mean-shift clustering method is appealing to practitioners and has been applied in many research and application areas, the underlying process is not well understood since the initial probability density function undergoes continuous nonlinear transformations. Chen (1995) pointed out that it is difficult to see where the mean-shift method leads to since all data points are moving simultaneously. The same statement is also true for some other non-parametric dynamic clustering methods. Although local properties of these algorithms are intuitively clear, the emerging properties have not been   well understood or established in the literature.
Although these methods have been   empirically validated,  a full understanding of the dynamic behaviors is still need to be established. A major difficulty to understand  the theoretical (dynamical) properties arises due to the complex spatiotemporal behavior of the probability density function driven by the aforementioned  nonlinear transformations in these dynamic clustering settings.

In this article, we initiate a dynamical system point of view for dynamical clustering algorithms by developing an analytical framework for dynamic clustering which prescribes the spatiotemporal evolution of the initial probability density function according to the rules imposed by the men-shift    clustering algorithm. The emerging properties derived from this proposed framework provide some useful criteria to evaluate the convergence and reliability for general class of mean-shift clustering methods. Our effort identifies the limitation of existing mean-shift clustering algorithms, and suggest potential improvements based on theoretical considerations.

The proposed framework is derived by modeling the successive nonlinear transformations of the underlying probability density function using the general  conservation law that characterizes and constrains the spatiotemporal evolution of the initial probability density function. This derived  differential form based on the  conservation law of the dynamic clustering scheme turns out to be second order partial differential equations (PDE) that is anti-diffusive in nature. Analytic study of this differential form enables us to conclude that a broad class of unsupervised mean-shift clustering algorithms can only converge for normal densities with independent structures.  This anti-diffusion has a Lyapunov function (an energy function that is non-increasing with respect to time) and the solution of such an anti-diffusion equation is uniquely determined only backward in time.  Kassoek (1980) provides a through and detailed theoretical analysis on the instability of one dimensional anti-diffusion phenomenon and concluded that the process is not stable in general.
We showed that a correct convergence can only happen when the underlying probability distribution has been successfully transformed
  to a multivariate normal distribution with no dependent structure. Due to the chaotic and unpredictable nature
  of the unsupervised mean shift algorithm, there is no guarantee that this will take place.  We then conclude that supervised clustering should be preferred in order to ensure a general and correct convergence. As a universally effective supervising function does not exist, a supervision  function may have to be chosen judiciously for each specific application. Our dynamic system point of view towards dynamic clustering is potentially useful  to choose a supervision function.

The rest of this paper is organized as follows. We present the analytical framework for dynamic clustering in Section 2.
Section 3 presents our theoretical analysis of the unsupervised dynamic clustering  and establishes the convergence of a supervised dynamic clustering. We conclude with a brief discussion
in Section 4.

\section{PDE Framework for Dynamic Clustering}

\subsection{Mean Shift Clustering and Statistical Mechanics}
In clustering analysis, a data set is viewed as a sample from a underling probability density function.
  For a particular sample ordered in a particular fashion, each observation is then fixed.
In dynamic clusterings such as mean-shift algorithm, however, data points are no longer static.
  Consequently, each data point can be viewed as a particle under a gravitational field or an autonomous agent governed by certain laws of attractions.
  To describe the location, speed and   direction of all the movements of such a complex system, it would require
  many variables or parameters to do so. Furthermore, such a full description of a complex system might not be necessary
  or possible for the purpose of clustering, we are only concerned about emerging patterns at the macroscopic level due to the self-organization of the data points in reaction the law of attraction imposed by any dynamic clustering method.

  In order to study the mean shift clustering algorithm applied to large number of data points, one can employ the analytical frameworks from statistical mechanics.
When studying collective behavior of small particles suspended in a stationary liquid, Einstein (1956) presented a
 a well known comprehensive framework which utilizes
 a differential equation framework for diffusion which is based on the underlying probability distribution. Since then, it is now the standard analytical framework for statistical mechanics. We employ exact the same framework except that the
 clustering process is the exactly the opposite of the diffusion in which data points are spreading out.
 Ellis (1985) provide a measure-theoretical justification of using probability distribution
 to study the macroscopic properties of a system with a large number of data points.
From an asymptotic point of view,    the underlying probability distribution can be estimated by an empirical or kernel density estimation with an arbitrarily   accuracy for a large sample. Detailed description and discussions of theoretical properties of kernel density estimation can be found in Simonoff (1996). In other words, large data points can be then be represented collectively by a continuous probability density function. When a large sample is available, the underlying probability distribution can then be represented by its empirical or kernel density function.

Since the data points are constantly moving,  the underlying continuous probability density function together with their empirical or kernel density  estimates undergoes a transformation process, which may could be described by examining the evolution of densities. The density is thus a special function depending on the spatiotemporal variable $(x,t)$ from the following set
\begin{equation}
{\cal E}=\{ f_t\; | f({\boldsymbol x}; t)\geq 0, \;\;\int f({\boldsymbol x}; t)d{\boldsymbol x} =1,\;\;t\in N,
{\boldsymbol x}\in R^n\}.
\end{equation}
 At the macroscopic level, instead of modelling the individual movement of a particular data point, we examine the pattern induced by the  transformations of the underlying probability distribution. A  dynamical clustering algorithm aims to identify the hidden probability distribution (backward in time) from which the given data distribution pattern is derived from the gravitattion/attraction law that governs the movements of  data
during the transformation  in time and space.

\vskip 0.05in

\subsection{Conservation law for clustering}

 In the dynamic clustering framework for the mean-shift algorithm, we assume that the total number of data points remain constant. If the algorithms do not delete or merge data points from the clustering process,  conservation law is then perfectly applicable so that the rate of change of the total number of particles contained in a fixed volume is equal to the influx of particles or data points passing through the boundary.

To illustrate our main point, we first consider the simple one dimensional case. Denote the one dimensional influx of data points by $ q (x; t)$ and the probability density by $f(x; t)$ at the spatial location $x$ and time $t$. We then have
\begin{equation}
 q (x; t)= u(x; t) \times f(x; t),
 \end{equation}
where   $u(x;t)$ is the speed of particles at location $x$ and time $t$.

In an unsurprised dynamic clustering, a constant flow of data points passes through an arbitrary small interval using the laws of attractions is characterized by the speed $u(x,t)$ that will be specified in later sections. Data  points  are assumed to be incompressible, and hence a standard argument in fluid dynamics in one-dimension space yields the {\it conservation law} given by
 \begin{equation}
 \frac{d\;f(x; t)}{dt} \;+\; \frac{d\;q(x; t) }{dx}=0.
 \end{equation}
This conservation law characterizes the functional connection between the probability density function and the influx function of
data points at a given spatial location and time. This fundamental view towards dynamic clustering provides the basis to establish an
analytical framework to describe the spatiotemporal evolution of the probability density function.

\subsection{The General PDE Framework }

 We now present the general differential form for high dimensional data. Denote an influx vector by ${\boldsymbol q}({\boldsymbol x}; t)$ and the probability density function  by $f({\boldsymbol x}; t)$.
 We then have
\begin{equation}
\frac{d}{dt} \int\limits_V f({\boldsymbol x}; t)\; dV = - \int\limits_S ({\boldsymbol q} . {\boldsymbol n} ) \;dS,
\end{equation}
where $dV$ is the volume element, $dS$ is the surface element of the boundary surface $S$, and
${\boldsymbol n}$ denotes the outward unit normal vector to $S$ with right-hand side measures the {\it outward}
influx indicated by the minus sign.

On applying the Gauss divergence theorem and taking $d/dt$ inside of the integral on the left hand side, we then have
\begin{equation}
\int\limits_V \left ( \frac{\partial f({\boldsymbol x}, t)}{\partial t} + {\nabla}  \; {\boldsymbol q}({\boldsymbol x}, t)\right ) dV =0,
\end{equation}
where $\nabla$ is the divergence operator given  by
 \begin{equation}
{\nabla} {\boldsymbol q}({\boldsymbol x}, t) \;=\; \sum\limits_{i=1}^n \frac{\partial^2 {\boldsymbol q}_i({\boldsymbol x}, t)}{\partial x_i},
\end{equation}
and ${\boldsymbol q}_i$'s are components of the vector ${\boldsymbol q}({\boldsymbol x}, t)$.

Since the result is valid for any arbitrary volume $V$, the integrand must be zero if it is continuous.
The differential form of the general conservation law is then given by
\begin{equation}\label{CN1}
 \frac{\partial f({\boldsymbol x}, t)}{\partial t} + {\nabla}\;   {\boldsymbol q}({\boldsymbol x}, t)\;=\;0,
\end{equation}
where $ {\boldsymbol q} (x; t)= {\boldsymbol u}(x; t) \times {\boldsymbol f}(x; t)$. We refer the detailed derivations and discussions of conservation laws and associated differential forms to   Debneth (2004).

For supervised dynamic clustering, the trajectories of data points will also be influenced or dictated by the introduction of certain
 supervision. This is equivalent to imposing a source or sink function in the dynamic process so data points will be
{\it absorbed} into a given domain. Denote such a supervision function by $\psi({\boldsymbol x}, t)$. A
 similar argument leads to the following conservation law with a sink function
\begin{equation}
 \frac{\partial f({\boldsymbol x}, t)}{\partial t} + {\nabla} q({\boldsymbol x}, t)\;=\; \psi({\boldsymbol x}, t).
\end{equation}

This framework is applicable to many dynamic clustering processes as long as data points are kept  in the partition process.
This framework gives us the foundation in the subsequent analytic analysis of emerging (backward in time) properties of dynamic clustering processes, including those unsupervised dynamic clustering approaches such as the mean-shift method.

\section{Properties of  Mean Shift Algorithm  }

\subsection{Unsupervised Mean Shift Clustering}
In unsupervised mean shift clustering methods, the movements of data points depend on the functional connection between   the current probability density function and its gradient or first order derivative.
The movements of many of its variant clustering methods are often governed by a law such that a data point will move to the center along more or less the direction of the gradient adjusted by the value of the current density function at the point of interest.
This can be formulated  mathematically as
\begin{equation} \label{assumption}
u({\boldsymbol x}; t) = a^2 \frac{\nabla f({\boldsymbol x}; t)} { f({\boldsymbol x}; t)}.
\end{equation}

Following the argument in Cheng(1995), one can show  that   the mean-shift algorithm indeed belongs to this category. All other varitions or improved versions based on the mean-shift method therefore are embraced by this category as well. A more general formulation for the movements in the traditional mean-shift method was discussed in Wang et al. (2007b) ,  in which data points move to the local center given by the conditional mean:
\begin{equation}
 {\boldsymbol x}^{k+1} = \frac{ \int_{B({\boldsymbol x}^k, d)}  \; {\boldsymbol t}\; f({\boldsymbol t})\; d \boldsymbol t}{\int_{B({\boldsymbol x}^k, d)} \;   f({\boldsymbol t})\; d \boldsymbol t},
\end{equation}
where $ B({\boldsymbol x}^k, d)$ is a neighborhood with the center located at ${\boldsymbol x}^k$ and the radius $d$.
Wang et al. (2007b) showed that, for any $\alpha>0$ and $d$ such that  $ \int_{B({\boldsymbol x}^k, d)}  \;   f({\boldsymbol t}) d {\boldsymbol t }=\alpha$, we have
\begin{equation}
{\boldsymbol x}^{k+1} = {\boldsymbol x}^{k} + \frac{n\;d^2}{n+2}\; \frac{\nabla f({\boldsymbol x^k}; t)} { f({\boldsymbol x^k}; t)}
+O(d^3).
\end{equation}
This justifies the assumption  (\ref{assumption}).

The gradient component forces each data point to optimize its trajectory to seek a local mode or cluster center which is known as the {\it mode seeking} property. The movement is also proportional to the reciprocal value of the current probability density function.
 This implies that data points in sparsely populated areas will travel longer distances when compared with that of data points in densely populated areas even if the gradient functions assume the same value at these two different locations.


\subsubsection{Anti-diffusion and Convergence for One-Dimensional Case}

Combining eqn (8) with the assumption described by eqn(\ref{assumption}), we obtain the corresponding differential form
as follows
\begin{equation}\label{1dCase}
 \frac{\partial }{\partial t} \;f(x; t) = - a^2\; \frac{d^2f(x; t)}{dx^2},
\end{equation}
where $a>0$ is a constant, and this equation has boundary  condition $f(x,0)=\phi_0(x)$, the initial probability density function.

This is a one-dimensional anti-diffusion equation, see Kaashoek (1980), in comparison with classical and popular diffusion where data points move from the region of high density to region of lower density. We now present the exact analytical solution to this differential equation.

\begin{theorem} \label{thm1}
Under the assumption (\ref{assumption}), the  one-dimensional anti-diffusion equation has  one unique
solution and takes the following form
\begin{equation}
f(x; t) \;\;=\;\;\frac{1}{\sqrt{-4 a^2 \pi t  } }
\int\limits_{-\infty}^{\infty} \phi_0(\xi) \; e^{ -\frac{(\xi - x)^2}{-  4 a^2 t} }\; d\xi, \;\;\;t\leq 0,
\end{equation}
where $f_0(x) = \phi_0(x)$, the initial probability density function.
\end{theorem}

{\bf  Proof:} This can be achieved by considering the diffusion equation $u(x, -t)$ for forward time and applying well-known results in linear reaction-diffusion equations. Here we use the classical transformation to give the detailed derivation. Consider the Fourier transformation of $f(x; t)$:
\begin{eqnarray*}
F(\omega; t) = \frac{1}{\sqrt{ 2\pi} }   \int\limits_{-\infty}^{\infty}
f(x, t) e^{i\omega x} dx;\;\;
f(x; t)=\frac{1}{\sqrt{ 2\pi} } \int\limits_{-\infty}^{\infty}
F(\omega, t) e^{-i\omega x} d\omega.\\
\end{eqnarray*}

Integration by parts yields
\begin{eqnarray*}
 \int\limits_{-\infty}^{\infty} \frac{\partial}{\partial t} f(x; t) \;e^{i\omega x}\; dx
 = \frac{\partial}{\partial t} \big ( \int\limits_{-\infty}^{\infty}  f(x; t) \;e^{i\omega x}\; dx \big )=\frac{\partial}{\partial t} F(\omega; t).
\end{eqnarray*}
and
\begin{eqnarray*}
 \int\limits_{-\infty}^{\infty} f_{xx}(x; t) \;e^{i\omega x}\; dx=
-   \omega^2 F(\omega, t).
 \end{eqnarray*}
It then follows from equation (\ref{1dCase}) that
\begin{equation}\label{1dCaseFT}
\frac{\partial}{\partial t} F(\omega; t)  - a^2  \omega^2 F(\omega, t)\;=\;0.
\end{equation}
The initial boundary condition also gives rise to
\begin{equation}
  \Phi_0(\omega) =F(\omega; 0) = \frac{1}{\sqrt{ 2\pi} } \int\limits_{-\infty}^{\infty}     \phi_0(x) e^{i\omega x} dx.
\end{equation}
Consequently, the equation (\ref{1dCaseFT}) has the solution
\begin{equation}
F(\omega; t)  = \Phi_0(\omega) e^{a^2 \omega^2 t},
\end{equation}
from which it follows that
\begin{equation}
f(x; t)\;=\;\frac{1}{\sqrt{2\pi}}\int\limits_{-\infty}^{\infty}     \Phi_0(\omega) e^{a^2 \omega^2 t - i \omega x}
d\omega.
\end{equation}
Note that the domain of convergence for the above integral  is $(-\infty, 0)$.

We then have
\begin{eqnarray*}
f(x; t)\;&=&\;\frac{1}{2\pi}\int\limits_{-\infty}^{\infty}
\left (
 \int\limits_{-\infty}^{\infty}     \phi_0(\xi) e^{i\omega \xi} d\xi\;\right)
 e^{a^2 \omega^2 t - i \omega x} d\omega\\
 &=& \frac{1}{2\pi}\int\limits_{-\infty}^{\infty}  \phi_0(\xi)
\left (
 \int\limits_{-\infty}^{\infty}    e^{  a^2 \omega^2 t + i \omega (\xi- x) } d\omega \right)
  d\xi.\\
  \;&=&\;  \frac{1}{2\pi}\int\limits_{-\infty}^{\infty}  \phi_0(\xi)
\left (
 \sqrt{ \frac{\pi}{a^2 (-t) } } e^{ (\xi-x)^2/4a^2t }\right)
  d\xi.\\
    \;&=&\;\frac{1}{\sqrt{-4 a^2 \pi t  } }
\int\limits_{-\infty}^{\infty} \phi_0(\xi) \; e^{ -\frac{(\xi - x)^2}{-  4 a^2 t} }\; d\xi, \;\;\;t\leq
0.\;\;\diamond
\end{eqnarray*}

The fact that the solution is specified uniquely only for $t\leq 0$ implies that the evolution of densities  produces  deterministic causal events.
Given the current status, there is only one unique
 process or path in the function space that  led to  what has occurred.
This conceptional observation about unsupervised clustering leads naturally to the following convergence result for dynamic clustering.

\begin{theorem} \label{thm2}
 Under the assumption (\ref{assumption}), we have
 \begin{itemize}
 \item[(i)] the convergence to a location $\mu_0$ of the clustering algorithm can only occur  at $t=0$ for normal densities with mean $\mu_0$ and variance;
proportional to $a^2$.
\item[(ii)] the first order derivative of the variance with respect to time is given by
\begin{equation}
\frac{d\sigma_t^2}{dt} = - 2 a^2,
\end{equation}
where $\sigma_t^2$ denotes the variance of the normal density at time $t$;
\item[(iii)] the converging speed of a data point at a  location $x$ at time $t$ is
\begin{equation}
u(x; t) = \frac{   x - \mu_0} {2 a^2 (-t)},\;\;\;t\leq 0.
\end{equation}
\end{itemize}
\end{theorem}

{\bf Proof:} If the convergence at time $t=0$ at a location $\mu_0$, this implies that
$f_0=\delta(x- {\bf\mu}_0)$. By Theorem \ref{thm1}, it then follows that
\begin{equation}
f(x; t)=\frac{1}{\sqrt{-4 a^2 \pi t  } }
 \;\;e^{ -\frac{(x - \mu_0)^2}{-  4 a^2 t} }, \;\;\;t\leq 0.
\end{equation}
The variance takes the form $2a^2(-t)$. Therefore, $\frac{d\sigma_t^2}{dt} = - 2 a^2.$
The rest of the result follow immediately from the assumption.$\diamond$

\smallskip
The conclusion of this theorem shows that, for one-dimensional data, a convergence to a single point can only occur for a normal density.
The spatial variation of normal densities at different time point depends on the magnitude of contraction.  In next section, we will prove that the same result holds true for higher-dimensional spaces when there are multiple cluster centers.

 \subsubsection{Convergence in Multi-dimensional Space without Supervision}

 Under the assumption (\ref{assumption}), it then follows  that
 \begin{equation}
 {\boldsymbol q}({\boldsymbol x}, t) = u({\boldsymbol x}, t)\;\; f({\boldsymbol x}, t) = a^2 {\nabla} f({\boldsymbol x}, t).
 \end{equation}
Consequently,   equation (\ref{CN1}) now becomes
\begin{equation}
\frac{\partial f({\boldsymbol x}, t)}{\partial t} + a^2\; \nabla^2 f({\boldsymbol x}, t)\; =\;0,
\end{equation}
where
the Laplacian of $f$,
$\nabla^2 f = \sum\limits_{i=1}^n  \partial ^2 f/\partial x_i^2$,  and  the boundary
condition  is given by $f({\boldsymbol x}, 0) = f_0({\boldsymbol x}).$

The phenomenon for the one dimensional case can be generalized to the multi-dimensional case when there are
multiple cluster centers.

\begin{theorem}
Under the assumption (\ref{assumption}), the anti-diffusion equation
\begin{equation}\label{eqn-thm3}
\frac{\partial f({\boldsymbol x}, t)}{\partial t} + a^2\; \nabla^2 f({\boldsymbol x}, t)\; =\;0,
\end{equation}
with the boundary condition $f|_{t=0}=\phi_0$  has the solution
 \begin{equation}
         f({\boldsymbol x}, t) \;=\; (-4a^2 t)^{-n/2} \;
         \int\limits_{(\boldsymbol \eta )}
         \phi_0({\boldsymbol \eta}) \;
         e^{- \frac{(\boldsymbol \eta - \boldsymbol x)^2 } { -4 a^2 t}  }\; d{\boldsymbol \eta},
\end{equation}
where $   (\boldsymbol \eta - \boldsymbol x)^2 = \sum\limits_{i=1}^n
(\eta_i - x_i)^2.$
\end{theorem}

{\bf Proof:} The dynamic shrinking or clustering can be characterized as
\begin{equation} \label{GN1}
\frac{\partial f({\boldsymbol x}, t)}{\partial t} + a^2\; \nabla^2 f({\boldsymbol x}, t)\; =\;0,
\end{equation}
with boundary condition $f({\boldsymbol x}, 0) = \phi_0({\boldsymbol x})$, a probability density function.

Denote the $n$-dimensional Fourier transformation by
\begin{equation}
F_n({\boldsymbol s}; t) =  \; \left( \frac{1}{\sqrt{2\pi}}\right )^n \int\limits_{-\infty}^{\infty} \int\limits_{-\infty}^{\infty}\dots
\int\limits_{-\infty}^{\infty}
f({\boldsymbol x}, t) e^{i \;{\boldsymbol s} \cdot {\boldsymbol x}}\; dx_1 dx_2 \dots dx_n,
\end{equation}
where $ {\boldsymbol s}\cdot{\boldsymbol x}= \sum\limits_{i=1}^n s_i\; x_i$. Applying this Fourier transformation on both sides of equation  (\ref{GN1}), we  have
\begin{equation}
      \partial F_n({\boldsymbol s},  t)/\partial t\;+\; a^2\;|| {\boldsymbol s}|| F_n({\boldsymbol s}, t) =0
\end{equation}
where   $||{\boldsymbol s}||= \sum\limits_{i=1}^n s_i^2$. The solution is given by
\begin{equation}
        F_n({\boldsymbol s}, t) \;=\; F_n({\boldsymbol s}, 0)\; e^{-a^2\;||{\boldsymbol s}||\;t}.
\end{equation}
where
\begin{equation}
F_n({\boldsymbol s}, 0)= \; \left( \frac{1}{\sqrt{2\pi}}\right )^n  \int\limits_{(\boldsymbol x)}  \; \phi_0({\boldsymbol x}) \; e^{i \;{\boldsymbol s} \cdot
{\boldsymbol x}}\; d{\boldsymbol x}.
\end{equation}
Using inverse Fourier transformation, we get
\begin{equation}
f({\boldsymbol x}, t) \;=\; \left( \frac{1}{\sqrt{2\pi}}\right )^n
  \int\limits_{(\boldsymbol s)}
       F_n({\boldsymbol s}, 0)\; e^{-a^2\;||{\boldsymbol s}||\;t - i\; {\boldsymbol s}\cdot{\boldsymbol x}}
       d{\boldsymbol s},
\end{equation}
therefore
\begin{equation}
 f({\boldsymbol x}, t) \;=\;  \left( \frac{1}{2\pi}\right )^{n} \int\limits_{(\boldsymbol s)} \left  (   \int\limits_{(\boldsymbol \eta)}
\; \phi_0({\boldsymbol \eta}) \;\; e^{i \;{\boldsymbol s} \cdot {\boldsymbol \eta}} \;
\; d{\boldsymbol \eta}  \right )\;
e^{-a^2\;||{\boldsymbol s}||\;t - i\; {\boldsymbol s}\cdot{\boldsymbol x}}
 d {\boldsymbol s}
\end{equation}
where ${\boldsymbol \eta}=(\eta_1, \eta_2, \cdots, \eta_n)$. By rearranging the order of integration and simplifying the same way in Theorem 3.1, we then have
\begin{equation}
         f({\boldsymbol x}, t) \;=\; (-4a^2 t\pi)^{-n/2} \;
         \int\limits_{(\boldsymbol \eta )}
         \phi_0({\boldsymbol \eta}) \;
         e^{- \frac{(\boldsymbol \eta - \boldsymbol x)^2 } { -4 a^2 t}  }\; d{\boldsymbol \eta},
\end{equation}
where $   (\boldsymbol \eta - \boldsymbol x)^2 = \sum\limits_{i=1}^n
(\eta_i - x_i)^2.$ $\diamond$

The analytic form allows us to retract the probability density function in the past and determine the convergence from a given initial probability density function. We can now show that the family of probability density functions that guarantee the convergence to distinct multiple cluster centres must be a multivariate normal distribution with independent correlation structures.

\begin{theorem} \label{thm4}
Under the assumption (\ref{assumption}), a dynamic shrinking or clustering  converges  to  $m$
distinct cluster centers if and only if the
density function is a mixture of normal distribution with equal variances, {\it i.e.}
\begin{equation}
f({\boldsymbol x}, t) = \sum\limits_{i=1}^m \lambda_i \;\phi_i, \;\;\;\;t<0.
\end{equation}
where $\phi_i$ is the normal density function with mean ${\boldsymbol\mu}_i$ and variance $-2a^2 t$.
\end{theorem}

{\bf Proof:} If the dynamic process converges to a finite number of focal points, {\it i.e.},
\begin{equation}
\phi_0(\boldsymbol \eta) = \sum\limits_{j=1}^m
\lambda_j \; \delta( \boldsymbol \eta - \boldsymbol \mu_j ),  \;\;\;\lambda_j\geq 0\;\;\;and\;\;\sum\limits_{j=1}^m
\lambda_j=1,
\end{equation}
where $ \boldsymbol \mu_j=(\mu_{j1}, \mu_{j2}, \cdots, \mu_{jn}),$ and $  \delta( \boldsymbol \eta - \boldsymbol
\mu_j )=
\prod\limits_{i=1}^n
\delta(\eta_j - \mu_{ji})$, then
\begin{equation}
 f({\boldsymbol x}, t) \;=\;   \sum\limits_{j=1}^m
 \lambda_j
 \left(\frac{1}{2\pi \sigma_t^2} \right)^{-n/2}
 e^{-\frac{( \boldsymbol x - \boldsymbol \mu_j)^2}{2\sigma_t^2} },
\end{equation}
where $\sigma_t^2 = - 2 a^2 t$, $t<0$.  $\;\;\;\diamond$

This theorem also implies that  the contraction rates  of dynamic  shrinking or clustering  must be homogenous in all directions at some time point of the clustering process to ensure
  a correct convergence. A persistent heterogeneous contraction pattern therefore implies a non-convergence.

\subsection{Instability of Unsupervised Dynamic Shrinking}

Despite the past success in applications of mean shift algorithm and its  intuitively appealing nature,
our theoretical analysis has shown that intrinsically this type of dynamics clustering algorithm actually
might not be able to converge correctly unless it can successively transform them
into  independent normal densities.

The instability of anti-diffusion for one one dimensional case  has been   established   in the literature,
see
Kaashoek (1980).   However, to the best of our knowledge, the
 results for higher dimensions have not been established.
In order to understand  more precisely about  the   instability of mean shift method for higher dimensions, we  now examine the temporal evolution of the system  using a quantity called an energy function or a Lyapunov function. This function has been widely used in dynamical systems and partial differential equations to describe the decay or growth of the system's energy. Detailed discussions can be found in Sastry (1999). In our case, this is a functional, for  a given density function $f$, of the following form:
\begin{equation}
H(f) =    - \int f({\boldsymbol x})\;\log f({\boldsymbol x})\; d{\boldsymbol x}, \end{equation}
In information theory, this is also cased the entropy, a measure of uncertainty, and for a normal distribution with mean $\mu$ and
variance $\sigma^2$, this entropy  is $\frac{1}{2} \log(2\pi \sigma^2 + 1)$.

This is called a Lyapunov functional due to the following property:
\begin{theorem}
Assume that the conservation law  (\ref{assumption}) holds.
If
\begin{equation}
\lim\limits_{x_i\rightarrow\infty}  \log f({\boldsymbol x})\;\;\frac{\partial f({\boldsymbol x}) }{\partial x_i}
=0, \;\;i=1,2,\cdots, m,
\end{equation}
 then
\begin{equation}
  \frac{d  H(f_t)}{dt} <0.
\end{equation}
\end{theorem}
{\bf Proof:}
Consider the first order derivative of $H(f_t)$ with respect to $t$. We then have
\begin{eqnarray*}
    \frac{d  H(f_t)}{dt}
     &= & - \int_{-\infty}^\infty
    \frac{\partial}{\partial t} \big (\; f({\boldsymbol x}) \log f({\boldsymbol x}) \;\big ) d{\boldsymbol x}   \\
    &=&  - \int_{-\infty}^\infty
    \left (  \frac{d f({\boldsymbol x})}{d t}\; \log f({\boldsymbol x}) + \frac{1}{f({\boldsymbol x})}\; f({\boldsymbol {\boldsymbol x}}) \; \frac{d f({\boldsymbol {\boldsymbol x}})}{d t}  \right ) d{\boldsymbol {\boldsymbol x}}     \\
   &=&  - \int_{-\infty}^\infty
   \left ( 1+ \log f({\boldsymbol x}) \right ) \frac{d f({\boldsymbol x})}{d t}  d{\boldsymbol x}\\
    &=&   \int_{-\infty}^\infty
     \left ( 1+ \log f(x) \right )   a^2 \left( \nabla^2 f({\boldsymbol x}, t)\right ) d{\boldsymbol x}\\
      &=&   \int_{-\infty}^\infty
     \left ( 1+ \log f(x) \right )   a^2 \left( \sum\limits_{i=1}^n
     \frac{\partial^2 f({\boldsymbol x}, t)}{\partial x_i^2} \right ) d{\boldsymbol x}\\
      &=&  - \int_{-\infty}^\infty
a^2 \frac{1}{f({\boldsymbol x})}\;\sum\limits_{i=1}^n \left( \frac{\partial f({\boldsymbol x})}{\partial x_i}\right )^2  d{\boldsymbol x}\;<0. \;\;\;\;\diamond
\end{eqnarray*}

So, in backward time, the Lyapunov function is non-decreasing at all times and hence, the dynamic clustering process is in complete violation of the second law of thermodynamics. In summary, we have thus observed that the dynamic shrinking and clustering does not correspond to a natural (physical) process and is unstable except for data with normal densities. To ensure a correct partion by using the mean shift algorithm,  a suitable intervention or supervision should be implemented.

\subsection{Convergence with Supervision}

Now we have established that the convergence of unsupervised shrinking or clustering can only be achieved through
a transformation to an
independent normal random variables with equal variances. However, this might not  be feasible due to
the instability described in previous section. A natural question arises: can this non-convergence
be overcome by some kind of supervision?  Mathematically, a possible formulation of this question is to impose the  so called {\it sink} or {\it force} function
into the PDE framework in the following form:
\begin{equation} \label{eqnSSS}
     \frac{\partial f({\boldsymbol x}, t)}{\partial t} + a^2\; \nabla^2 f({\boldsymbol x}, t)\; =\; \psi({\bf x},
     t),
\end{equation}
where $\psi$ is a continuous function.

We now show that a correct convergence can be established through non-normal densities with the help of supervision
function.
 \begin{theorem}
 Under the assumption (\ref{assumption}), we have
 \begin{itemize}
 \item[(i)] the PDE associated with supervised clustering has the following solution
  \begin{equation}
\begin{split}
f({\boldsymbol x}, t) &=
         (-4a^2 t\pi)^{-n/2} \;
         \int\limits_{(\boldsymbol \eta )}
         \phi_0({\boldsymbol \eta}) \;
         e^{- \frac{(\boldsymbol \eta - \boldsymbol x)^2 } { -4 a^2 t}  }\; d{\boldsymbol \eta}\\
         \;\;&+
\;\int^0_t\int_{({\boldsymbol \xi})}\;\psi({\boldsymbol \xi}, \tau)\;
[-4a^2 (t-\tau) \pi]^{-n/2}  \;\; e^{- \frac{(\boldsymbol x - \boldsymbol \eta)^2 } { -4 a^2 (t-\tau)}  }
d{\boldsymbol \xi}\;d\tau, \;\;\;t\leq 0.
\end{split}
\end{equation}
\item[(ii)] if the clustering process converges to $m$
distinct focal points, then
 \begin{equation}
\begin{split}
f({\boldsymbol x}, t) &=
\sum\limits_{j=1}^m
 \lambda_j
 \left(\frac{1}{2\pi \sigma_t^2} \right)^{-n/2}
 e^{-\frac{( \boldsymbol x - \boldsymbol \mu_j)^2}{2\sigma_t^2} }\\
 \;\;&+
\;\int^0_t\int_{({\boldsymbol \xi})}\;\psi({\boldsymbol \xi}, \tau)\;
[-4a^2 (t-\tau) \pi]^{-n/2}  \;\; e^{- \frac{(\boldsymbol x - \boldsymbol \eta)^2 } { -4 a^2 (t-\tau)}  }
d{\boldsymbol \xi}\;d\tau, \;\;\;t\leq 0.
\end{split}
\end{equation}
\end{itemize}
 \end{theorem}

 {\bf Proof:}  The general solution with the sink function in the PDE can be decomposed into two parts:
 \begin{equation}
 f({\boldsymbol x}, t) = g_1({\boldsymbol x}, t) + g_2({\boldsymbol x},t),
 \end{equation}
 where $g_1$ is the solution for the PDE (\ref{eqn-thm3}) with boundary condition
 $g_1({t=0})=\phi_0$ and $g_2$ satisfying the PDE (\ref{eqnSSS}) with the boundary condition
 $g_{t=0}^2=0.$ The function form of $g_1$ is given by Theorem 3. That is,
 \[
         g_1({\boldsymbol x}, t) \;=\; (-4a^2 t\pi)^{-n/2} \;
         \int\limits_{(\boldsymbol \eta )}
         \phi_0({\boldsymbol \eta}) \;
         e^{- \frac{(\boldsymbol \eta - \boldsymbol x)^2 } { -4 a^2 t}  }\; d{\boldsymbol \eta}, \;\;  (\boldsymbol \eta - \boldsymbol x)^2 = \sum\limits_{i=1}^n
(\eta_i - x_i)^2.
\]

To find $g_2$, we consider a nonhomogeneous differential equation of the form
\begin{equation}\label{G0}
L_{\boldsymbol x}\; u({\boldsymbol x}) = \psi({\boldsymbol x}, t),
\end{equation}
where
\begin{equation}
L_{\boldsymbol x}\; u({\boldsymbol x})= \frac{\partial u({\boldsymbol x}, t)}{\partial t} + a^2\; \nabla^2 u({\boldsymbol x}, t).
\end{equation}

The Green function $G({\boldsymbol x}, {\boldsymbol \xi})$ of this problem satisfies the equation
\begin{equation}
L_{\boldsymbol x}\; G({\boldsymbol x}, {\boldsymbol \xi}) = \delta(\boldsymbol x-{\boldsymbol \xi})
\delta(t-\tau), \;\; G_{t=0}=0.
\end{equation}
The solution for the partial differential equation (\ref{G0}) is then given by
 \begin{equation}
 u({\boldsymbol x})\;=\;\int^0_t\int_{({\boldsymbol \xi})}\;\psi({\boldsymbol \xi}, \tau)\;G({\boldsymbol x}, t;  {\boldsymbol \xi}, \tau)\; d{\boldsymbol \xi} d\tau,
 \end{equation}
where the Green function satisfying the following PDE
\begin{eqnarray*} \frac{\partial\; G({\boldsymbol x}, t)}{\partial t} + a^2\; \nabla^2 G({\boldsymbol x}, t)=0,\;\; G|_{t=\tau} = \delta({\boldsymbol x}- {\boldsymbol \xi}).
\end{eqnarray*}

By theorem 3 and replacing $t$ by $t-\tau$, the Green function is then given by
\begin{equation}
\begin{split}
         G({\boldsymbol x}, t) \;&=\; [-4a^2 (t-\tau)\pi]^{-n/2} \;
         \int\limits_{(\boldsymbol \eta )}
         \delta({\boldsymbol \eta}- {\boldsymbol \xi}) \;
         e^{- \frac{(\boldsymbol \eta - \boldsymbol x)^2 } { -4 a^2 (t-\tau)}  }\; d{\boldsymbol \eta}\\
        &= [-4a^2 (t-\tau) \pi]^{-n/2}  \;\; e^{- \frac{(\boldsymbol x - \boldsymbol \eta)^2 } { -4 a^2 (t-\tau)}  },
         \; \;\;\; (\boldsymbol x- \boldsymbol \eta)^2 = \sum\limits_{i=1}^n
( x_i - \eta_i)^2.
\end{split}
\end{equation}
where $   (\boldsymbol x- \boldsymbol \eta)^2 = \sum\limits_{i=1}^n
( x_i - \eta_i)^2.$

It then follows that
\begin{equation}
g_2({\boldsymbol x}, t) =
\;\int^0_t\int_{({\boldsymbol \xi})}\;\psi({\boldsymbol \xi}, \tau)\;
[-4a^2 (t-\tau) \pi]^{-n/2}  \;\; e^{- \frac{(\boldsymbol x - \boldsymbol \xi)^2 } { -4 a^2 (t-\tau)}  }
d{\boldsymbol \xi}\;d\tau, \;\;\;t\leq 0.
\end{equation}
Therefore,
\begin{equation}
\begin{split}
f({\boldsymbol x}, t) &=\frac{1}{M}
         (-4a^2 t\pi)^{-n/2} \;
         \int\limits_{(\boldsymbol \eta )}
         \phi_0({\boldsymbol \eta}) \;
         e^{- \frac{( \boldsymbol x- \boldsymbol \eta )^2 } { -4 a^2 t}  }\; d{\boldsymbol \eta}\\
         \;\;&+ \frac{1}{M}
\;\int^0_t\int_{({\boldsymbol \xi})}\;\psi({\boldsymbol \xi}, \tau)\;
[-4a^2 (t-\tau) \pi]^{-n/2}  \;\; e^{- \frac{(\boldsymbol x - \boldsymbol \xi)^2 } { -4 a^2 (t-\tau)}  }
\;\;d{\boldsymbol \xi}\;d\tau, \;\;\;t\leq 0,
\end{split}
\end{equation}
where $M$ is the normalizing constant to ensure that $f({\boldsymbol x}, t)$ is a proper probability density function.
 $\diamond$

The fact that the original density function of the PDE is a function of the supervision function implies
 that a correct convergence   is dependent on the choice of the supervising function. The assertion of the theorem
 indicates that a universally effective supervising function might  not exist.
 A supervising function then must be chosen judiciously to ensure a correct convergence.
A self-adaptive learning algorithm will also require that the sink function be a functional of the current and historical densities, and this issue is left for future studies.

We remark that there are some dynamic shrinking and clustering algorithms that could be stable due to external sink functions.
This is due to the design of these algorithms to deal with the intrinsic violation of the conservation law.
One such an example is the crystallization processes as described in Teran and Bill (2010). It is
stable due to the fact that particles are accumulating and transformed into solid with zero speed due to the crystallization. How can this inspire a certain choice of  supervision for dynamic clustering remains to be an interesting topic for future studies.

\section{Discussion}
We aim to fill the critical gap in the literature between the reported success in data mining applications and the lack of convergence and stability analysis of mean shift type of dynamic clustering algorithms. We employ the conservation law from physics and establish the general partial differential equation framework that prescribes the spatiotemporal evolutions of dynamic clustering processes. We show that, in the absence of a sink function or supervision, mean shift clustering and its variations  may not result in a correct convergence in general unless the underlying probability distribution can be transformed to normal densities. The non-decreasing backward in time of the Lyapunov function and the anti-diffusion nature of these dynamic clustering algorithms render them universally highly unreliable without a proper supervision. As such, supervised mean shift clustering should be  preferred and the supervising function must be chosen carefully to ensure valid results. We have indicated how this supervision function should be incorporated into the anti-diffusion formulation. The   emerging properties of general supervised mean shift  clustering could be rich.

\vskip 0.1in

\parindent 0pt



\begin{thebibliography}{9}



\bibitem{Choi-Hall-1999}
Choi, E. and Hall, P. (1999). Data sharpening as a prelude to density estimation
{\it Biometrika}, {\bf 86}, 941-947.




\bibitem{Thomas-Cover-2006}
Cover, T.M. and Thomas, J.A. (2006) {\it Elements of Information Theory}. John Wiley\& Sons, Inc. New Jersy.



 \bibitem{Cheng-1995} Cheng, L. (1995) Mean shift, mode seeking, and clustering.
 {\it IEEE Transactions on Pattern Analysis and Machine Intelligence}, {\bf 17}, 790-799.


\bibitem{ComaniciuMeer:2002}
{Comaniciu, D.} and {Meer, P.} (2002).
 Mean shift: a robust approach toward feature space analysis.
 {\em IEEE transactions on pattern analysis and machine intelligence},
  24\penalty0 (5):\penalty0 603--619.



\bibitem{DEbnath:2004}
{Debnath, L. } (2004).
{\em  Nonlinear Differential Equations for Scientists and Engineers.}
Birkhauser, Boston.


\bibitem{Ellis:1985}
{Ellis, S. Richard} (1985).
{\em Entropy, Large Deviations and Statistical Mechanics.}
Springer, New York.


\bibitem{Einsterin:1956}
  Einstein, Albert (1956). Investigations on the Theory of the Brownian Movement. Dover.


\bibitem{FraleyRaftery:2002}
{Fraley, C.} and {Raftery, A. E.} (2002).
 Model-based clustering, discriminant analysis, and density
  estimation.
 {\em Journal of the American Statistical Association}, 97\penalty0
  (458):\penalty0 611--631.

\bibitem{Frigui:1999}
{Frigui, H.} and {Krishnapuram, R.} (1999).
 A robust competitive clustering algorithm with applications in
  computer vision.
 {\em IEEE Transactions on Pattern Analysis and Machine Intelligence},
  21\penalty0 (5):\penalty0 450--465.

\bibitem{FukunagaHostetler:1975}
{Fukunaga, K.} and {Hostetler, L. D.} (1975).
 The estimation of the gradient of a density function, with
  applications in pattern recognition.
 {\em IEEE Transactions on Information Theory}, 21:\penalty0 32--40.



\bibitem{Han:2006}
Han, J. and Kamber, M. (2006).
{\it Data Mining: Concepts and Techniques, 2nd edition}.
The Morgan Kaufmann Series in Data Management Systems.


\bibitem{Hartigan:1979}
{Hartigan, J. A.} and {Wong, M. A.} (1979).
 A k-means clustering algorithm.
 {\em Applied Statistics}, 28:\penalty0 100--108.

\bibitem{HastieEtAl:2001}
{Hastie, T.}, {Tibshirani, R.}, and {Friedman, J.} (2001).
 {\em The elements of statistical learning: data mining, inference,
  and prediction}.
 Springer-Verlag.

\bibitem{Gan:2007}
Gan, G.,   Ma, C. and   Wu, J. (2007).
Data Clustering: Theory, Algorithms, and Applications.
ASA-SIAM Series on Statistics and Applied Probability.

\bibitem{Kaashoek (1980)}
Kaashoek, J.F. (1980).
{\em  Modelling one dimensional pattern formation by anti-diffusion.}
Center for Mathematics and Computer Science, CWI Tract.

\bibitem{Kaufman:1990}
{Kaufman, L.} and {Rousseeuw, P. J.} (1990).
 {\em Finding Groups in Data: An Introduction to Cluster Analysis}.
 Wiley, New York.

\bibitem{Kundu:1999}
{Kundu, S.} (1999).
 Gravitational clustering: a new approach based on the spatial
  distribution of the points.
 {\em Pattern Recognition}, 32:\penalty0 1149--1160.


\bibitem{MacQueen:1967}
{MacQueen, J. B.} (1967).
 Some methods for classification and analysis of multivariate
  observations.
 {\em Proceedings of 5th Berkeley Symposium on Mathematical
  Statistics and Probability, 1}, pages 281--297. Berkeley, Calif: University
  of California Press.


\bibitem{Sato:2000}
{Sato, Y.} (2000).
 An autonomous clustering technique.
 In {Kiers, A. L. Henk}, {Rasson, Jean-Paul}, {Groenen, Patrick J.
  E.}, and {Schader, Martin}, editors, {\em Data analysis, classification, and
  related methods}. Springer.



\bibitem{SAS-1999} Sastry, S. (1999).
Nonlinear Systems: Analysis, Stability and Control.
   Springer-Verlag, New York.

 \bibitem{Shi-1995} Shi, Y.;   Song, Y.;   Zhang, A. (2005)
A shrinking-based clustering approach for multidimensional data.
{\it  IEEE Transactions on Knowledge and Data Engineering}, {\bf 17}, 1389-1403.

\bibitem{Shi-1995} Simonoff, J.S. (1996).
Smoothing Methods in Statistics.
   Springer-Verlag, New York.


 \bibitem{Teran-Bill-1995}
 Teran, A. V. and Bill, A. (2010).
Time-evolution of grain size distributions in random nucleation and growth
crystallization processes.
{\it  Physics Review}, {\bf 81}, 19.

\bibitem{Tibshirani:2000}
{Tibshirani, R.}, {Walther, G.}, and {Hastie, T.} (2000).
 Estimating the number of clusters in a dataset via the gap statistic.
 {\it Technical Report 208}, Dept. of Statistics, Stanford University.

 \bibitem{Virmajoki-2002}
Virmajoki, O.;   Franti, P.;   Kaukoranta, T. (2002) Iterative shrinking method for generating clustering.
{\it Proceedings of the International Conference on Image Processing,}, {\bf 2}, 685-688.

\bibitem{Wang:2001}
{Wang, J. H.} and {Rau, J. D.} (2001).
 {\sc VQ}-agglomeration: A novel approach to clustering.
 {\em IEE Proceedings-Vision, Image and Signal Processing},
  148\penalty0 (1):\penalty0 36--44.



\bibitem{Wang-2007}
Wang, X., Qiu, W. and Zamar, H. R. (2007a). CLUES: A non-parametric clustering method based on local shrinking.
Computational Statistics and Data Analysis,
{\bf 52}, 286-298.

\bibitem{Wang-2006}
Wang, X., Liang, D, Feng, X. and Ye, L. (2007b) A derivative-free optimization algorithm based on
conditional moments, Mathematical Analysis and Applications, {\bf 331},  1337-1360.



\bibitem{Woolfold-2007}
Woolfold, D. G. and Braun, W. J. (2006).
Convergent data sharpening for the identification and tracking of spatial temporal centers of lightning activity,
{\it Envirometrics},
{\bf 18},   461-479.


\bibitem{Wright:1977}
{Wright, W. E.} (1977).
 Gravitational clustering.
 {\em Pattern Recognition}, 9:\penalty0 151--166.

\bibitem{Zhung:1996}
{Zhung, X.}, {Huang, Y.}, {Palaniappan, K.}, and {Lee, J. S.} (1996).
 Gaussian mixture modeling, decomposition and applications.
 {\em IEEE Transactions on Signal Process}, 5:\penalty0 1293--1302.




\end{thebibliography}
\end{document}